\documentclass[%
 reprint,
 amsmath,amssymb,
 aps,
]{revtex4-1}

\usepackage{graphicx}
\usepackage{dcolumn}
\usepackage{bm}
\usepackage{booktabs}
\usepackage{subfigure}
\usepackage{float}
\usepackage{amsmath}
\usepackage[titletoc]{appendix}

\begin{document}
\preprint{APS/123-QED}

\title{Analytical study of mode degeneracy in non-Hermitian photonic crystals with TM-like polarization}

\author{Xuefan Yin$^1$}
\author{Yong Liang$^2$}
\author{Liangfu Ni$^1$}
\author{Zhixin Wang$^1$}
\author{Chao Peng$^1$}
\email{pengchao@pku.edu.cn}
\author{Zhengbin Li$^1$}
\affiliation{$^1$State Key Laboratory of Advanced Optical Communication Systems $\&$ Networks, Department of Electronics, Peking University, Beijing 100871, China\\
$^2$Institute of Quantum Electronics, ETH Zurich, 8093 Zurich, Switzerland}
\date{\today}

\begin{abstract}
We present a study of the mode degeneracy in non-Hermitian photonic crystals (PC) with TM-like polarization and $C_{4v}$ symmetry from the perspective of the coupled-wave theory (CWT). The CWT framework is extended to include TE-TM coupling terms which are critical for modeling the accidental triple degeneracy within non-Hermitian PC systems. We derive the  analytical form of the wave function and the condition of Dirac-like-cone dispersion when radiation loss is relatively small. We find that, similar to a real Dirac cone, the Dirac-like cone in non-Hermitian PCs possesses good linearity and isotropy, even with a ring of exceptional points (EPs) inevitably existing in the vicinity of the 2nd-order $\Gamma$ point. However, the Berry phase remains zero at the $\Gamma$ point, indicating the cone does not obey Dirac equation and is only a Dirac-\textit{like} cone. The topological modal interchange phenomenon and non-zero Berry phase of the EPs are also discussed.
\end{abstract}

\pacs{Valid PACS appear here}
\maketitle

\section{\label{sec:level1}INTRODUCTION}
Photonic crystals (PCs) have been attracted much attention in the past years owing to their great potential in manipulating light at wavelength scale. Various devices and applications, such as optical waveguide \cite{46,47}, filter \cite{35,43,44}, cavity \cite{42,45}, and laser \cite{36} have been demonstrated. Conventionally, the light is vertically confined within the PC slab because of the total internal reflection at dielectric boundaries, and hence, such closed and lossless systems can be described by Hermitian operators. Recently, some extraordinary phenomena have been observed within the radiation continuum of PC slabs where light is allowed to escape and transport energy away \cite{1,2,3,4}. 
Anomalously, narrow resonances can occur in the continuum. Such unique resonances have been interpreted as a photonic analogy of the bound states in the continuum (BICs) in quantum mechanics that were first proposed by von Neumann and Wigner \cite{41} and have been intensively studied in decades \cite{5,6,7,38}. Apparently, such PC systems are non-Hermitian since energy conservation no longer holds.

Owing to the analogy between quantum mechanics and electrodynamics, PCs possess photonic band that is similar to the electronic band of crystalline solids. It was recently found that, some accidental degeneracies can occur at the center of the Brillouin zone of a square \cite{14,15,16} or triangular \cite{8,9,10,11,12,13} PC lattice, leading to the observation of Dirac-like cone \cite{13} and various counter-intuitive transport properties such as zero-index metamaterials \cite{14,15,20,21} and ring of EPs \cite{22}. 

Previously, the 2D Hermitian PC systems were studied by using effective medium theory \cite{10,11}, $\vec{k} \cdot \vec{p}$ method \cite{25}, and tight-binding approximation \cite{16,20}. In these 2D PC systems, there is no radiation loss since the PCs are assumed to be infinitely thick. The triple degeneracy is found by continuously varying the radii of the dielectric pillar. 
However, radiation loss exists in the non-Hermitian PC slabs, leading to non-orthogonal eigenfunctions with complex eigenvalues. Therefore, we have to take into account more structural parameters to form the accidental degeneracy.
Meanwhile, intriguing phenomena such as EPs have been experimentally demonstrated \cite{22} which reveal some essential differences between Hermitian and non-Hermitian systems. As clarified by \citet{25} for Hermitian PC systems, the Dirac-like cone at Brillouin zone center with $C_{4v}$ symmetry has zero Berry phase, and hence it can not be mapped into the massless Dirac equation like the Dirac cone in Graphene. Consequently, such Dirac-like cone is not expected to give rise to extraordinary properties such as \textit{Zitterbewegung} \cite{9,24} and anti-localization. 
Yet, for non-Hermitian PC systems with Dirac-like cone, it is still unclear if such a claim holds.

In recent years, we have developed a comprehensive coupled-wave theory (CWT) framework capable of analytically modeling non-Hermitian PC systems \cite{26,27,28,29,30,31}. Since the radiation loss has been included in the analysis, CWT can be a promising tool to investigate the mode degeneracy in non-Hermitian systems. 
In this work, we extend the CWT framework to analytically study the complex band structure near the 2nd-order $\Gamma$ point of PC slabs with TM-like polarization and $C_{4v}$ symmetry. 
In contrast to our previous works, we include the TE-TM coupling terms which are crucial for the study of mode degeneracy within such non-Hermitian PC systems. 
Furthermore, we obtain the reduced coupling matrix as well as analytical wave functions, which provide analytical insights into the characteristics of band structure for non-Hermitian systems. 

The remainder of this paper is organized as follows. 
In Section \ref{sec:level2}, we derive the extended CWT formulation and the coupling matrix by including the important TE-TM coupling terms.
In Section \ref{sec:level3}, we first analytically derive the explicit condition of forming the triple degeneracy. Then, we obtain a reduced coupling matrix from the CWT equation to analytically study the linearity and isotropy of the band structure. 
It is found that, a ring of EPs inevitably appears around the 2nd-order $\Gamma$ point if the system is non-Hermitian. 
The topological properties of the triple degeneracy is discussed in Section \ref{sec:level4}. In Section \ref{sec:level5}, we conclude with our findings. 


\section{\label{sec:level2}Theory and Formulation}

\begin{figure}[htbp]
\centering
\includegraphics[width=8.5cm]{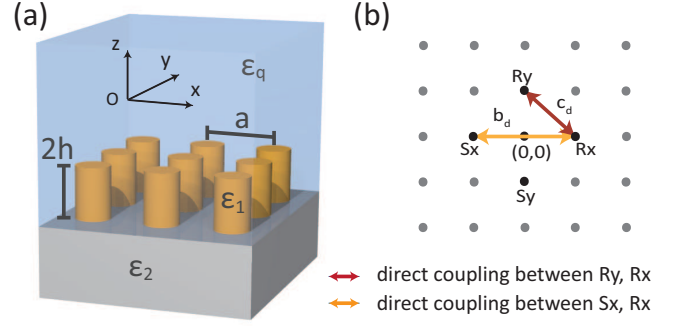}
\caption{(a) Schematic of a multilayered PC slab consisting of a square latticed pillars on the substrate. The PC pillars are embedded in refractive-index-matching liquid whose permittivity is denoted as $\varepsilon_{q}$. (b) Two typical direct-coupling paths (arrows) between the basic waves depicted in reciprocal space.}
\label{fig:1}
\end{figure}

In this section, we present the coupled-wave formulation for a two dimensional (2D) PC slab with periodicity in the $x$ and $y$  directions and multilayered structure in $z$ direction. As shown in Fig. \ref{fig:1}(a), the PC layer consists of a square lattice of circular-shaped pillars with two different materials (permittivities: $\varepsilon_1$ and $\varepsilon_q$). The PC pillar has a finite thickness $2h$ with lattice constant $a$ and filling factor $f$. Under TM-like polarization, the $\mathbf H$ component follows:
\begin{eqnarray}
\label{eq:1}
\nabla\times[\frac{1}{\varepsilon(\mathbf{r})}\nabla\times\mathbf{H}(\mathbf{r})]=k^2\mathbf{H}(\boldsymbol{r}).
\end{eqnarray}

According to the Bloch's theorem, For an off-$\Gamma$ reciprocal wave vector $\boldsymbol{k}_{\parallel} =k_x \hat{e_x}+k_y \hat{e_y}$, the components $H_{i}(z)$ can be expanded as $\sum_{m,n}H_{i,mn}(z)e(m,n)$ where $e(m,n)=\exp(-im_x\beta_0x-in_y\beta_0y),~i=x,y,z$, $\beta_0={2 \pi}/{a}$ and $m_x=m+k_x/\beta_0, n_y=n+k_y/\beta_0$. We also expand ${1}/{\varepsilon(\mathbf{r})}$ as $\kappa_{0}+\sum_{m\neq 0,n\neq 0}{\kappa}_{m n}e(m,n)$. For the PC layer, upper-clad and lower-clad, we denote $\kappa_0(z)$ as $\kappa_a=f/\varepsilon_1+(1-f)/\varepsilon_q$, $\kappa_b=1/\varepsilon_q$ and $\kappa_c=1/\varepsilon_2$, respectively.

For TM-like polarization that we focus on, the magnetic fields are given as $(H_x,H_y,H_z)$. The $H_z$ component was neglected in our previous works because of the transverse nature of the TM mode. However, for a realistic and non-Hermitian PC slab, $H_z = 0$ only rigorously holds at one particular cross section of the slab. The existence of $H_z$ component implies that the system contains TE components $\{E_x,E_y,H_z\}$, which will lead to some extra coupling paths that we denote as TE-TM couplings. Hence, the $H_z$ terms are included in the formulation to improve the accuracy (see Appendix A).

The set of coupled-wave equations that depicts the wave interactions within non-Hermitian PC slab is presented as (also see previous works \cite{26,27,31}):

\begin{widetext}
\begin{eqnarray}
\label{eq:2.a}
\mathcal{H}(n_y H_{x,m n}-m_x H_{y,m n})&=&\sum \limits_{m',n'\neq m,n}\kappa_{\substack{m-m'\\n-n'}}\{\Pi_p(H_{x,m' n'},H_{y,m' n'})+\Pi_s(H_{x,m' n'},H_{y,m' n'},H_{z,m' n'})\}\\
\label{eq:2.b}
\mathcal{H}(m_x H_{x,m n}+n_y H_{y,m n})+\mathcal{H_\delta}H_{z,mn}&=&\sum \limits_{m',n'\neq m,n}\kappa_{\substack{m-m'\\n-n'}}\{\Xi_p(H_{x,m' n'},H_{y,m' n'})+\Xi_s(H_{x,m' n'},H_{y,m' n'},H_{z,m' n'})\}\\
\label{eq:2.c}
\mathcal{H}_zH_{z,m n}&=&\sum \limits_{m',n'\neq m,n}\kappa_{\substack{m-m'\\n-n'}}\{\Upsilon_p(H_{x,m' n'},H_{y,m' n'},H_{z,m' n'})\}
\end{eqnarray}
\end{widetext}

where
\begin{eqnarray}
\label{eq:3}
\mathcal{H}&=&k^2+\kappa_{0}\frac{\partial^2}{\partial z^2}+\delta_K\frac{\partial}{\partial z}-\kappa_{0}(m_x^2+n_y^2)\beta_0^2\nonumber\\
\mathcal{H}_z&=&k^2+\kappa_{0}\frac{\partial^2}{\partial z^2}-\kappa_{0}(m_x^2+n_y^2)\beta_0^2\nonumber\\
\mathcal{H_{\delta}}&=&i\beta_0\delta_K(m_x^2+n_y^2)\nonumber
\end{eqnarray}
and $\delta_K=(\kappa_b-\kappa_a)\delta(z-h)+(\kappa_a-\kappa_c)\delta(z+h)$. The right hand side of Eqs.(\ref{eq:2.a})-(\ref{eq:2.c}) depicts the coupling paths between individual Bloch waves, which can be classified into two groups: the terms $\Pi_p,\Xi_p, \Upsilon_p$ with subscript $p$ describe the in-plane couplings that are induced by the planar permittivity modulation (the only coupling mechanism for TE-like modes); the terms $\Pi_s,\Xi_s$ with subscript $s$ represent the surface couplings owing to the independent boundary condition in TM-like polarization. By applying the boundary conditions appropriately (as discussed in Appendix \ref{App:A}), such coupled-wave equations can be solved.

Submitting Eqs.(\ref{eq:2.c}) to (\ref{eq:2.b}), we obtain the transverse constraint:
\begin{equation}
\label{eq:4}
i\beta_0(m_xH_{x,m n}+n_y H_{y,m n})=\frac{\partial}{\partial z}H_{z,m n}
\end{equation}
which is exactly the same as the divergence equation $\nabla\cdot\boldsymbol{H}=0$. For vertical homogeneous structure, or the case where the TE-TM coupling is sufficiently weak, Eq.(\ref{eq:4}) turns out to be a trivial form of $m_x H_{x,m n}+n_y H_{y,m n}=0$.

We choose a set of Bloch waves $\mathbf V$ as the basis to simplify Eqs.(\ref{eq:2.a})-(\ref{eq:2.c}). At the 2nd-order $\Gamma$ point, the modal energy is dominated by the Bloch waves with in-plane wave vectors satisfying ${m^2+n^2}=1$, which are referred to as basic waves. Additionally, we denote the Bloch waves ${m^2+n^2}>1$ and ${m^2+n^2}=0$ as high-order waves and radiative wave, respectively; these waves are excited by the basic waves.

For an arbitrary wave vector $\boldsymbol{k}_{\parallel}$ near the 2nd-order $\Gamma$ point, the basic waves follow:
\begin{eqnarray}
\label{eq:5}
n_y H_{x,m n} - m_x H_{y,m n} &=& \sqrt{m_x^2+n_y^2}V_{m n}\Theta_{0,m n}(z)
\end{eqnarray}
where $\Theta_{0,m n}, (m^2+n^2=1)$ are the vertical (out-of-plane) profile of the basic waves; and we denote $V_{m n}$ as $\{R_x,S_x,R_y,S_y\}$ by giving $(m,n)\in\{(1,0),(-1,0),(0,1),(0,-1)\}$. Since the TE-TM coupling terms are relatively weak, we assume that the basic waves are almost in TM polarization, and the trivial transverse constraint $m_x H_{x,m n}+n_y H_{y,m n} = 0$ still holds. We take the basic waves as the unperturbed basis and the couplings as perturbation. Hence, $\Theta_{0, mn}$ should satisfy
\begin{equation}
\label{eq:7}
\mathcal{H}(m,n)\Theta_{0,m n}(z)=0
\end{equation}

By solving Eqs.(\ref{eq:5})-(\ref{eq:7}) with trivial transverse constraint, the basic waves, including the vertical profile $\Theta_{0,m n}$ and the wavenumber $k_{0,m n}$, can be obtained. Besides, the high-order waves and radiative waves can be solved by using Green function method \cite{26,27,30}. Hence, we can rewrite the equations into an eigenvalue problem:
\begin{equation}
\label{eq:9}
k^2\mathbf{V}=(\boldsymbol{K}+\boldsymbol{C})\mathbf{V}=\boldsymbol{D}\mathbf{V}
\end{equation}
where $\mathbf{V}=[R_x~S_x~R_y~S_y]^T$, and matrix $\boldsymbol{K}$ is in a diagonal form of $\{k_{0,10}^2, k_{0,-10}^2, k_{0,01}^2, k_{0,0-1}^2\}$.

The coupling matrix $\boldsymbol{C}$ depicts the coupling strength between the individual basic waves, whereas the other orders of Bloch waves contribute as coupling paths. Here, the matrix $\boldsymbol{C}$ can be written as:
\begin{equation}
\label{eq:11}
\boldsymbol{C}=\boldsymbol{C}_{d}+\boldsymbol{C}_{rad}+\boldsymbol{C}_{h}+\boldsymbol{C}_{TE\text{-}TM}
\end{equation}
where $\boldsymbol{C}_{d}$, $\boldsymbol{C}_{rad}$, $\boldsymbol{C}_{h}$ and $\boldsymbol{C}_{TE\text{-}TM}$ correspond
to the direct couplings between basic waves [see Fig. \ref{fig:1}(b)], the couplings via radiative waves, couplings via high-order waves, and the TE-TM couplings, respectively. Since the basic waves as well as high-order waves are confined within the PC slab, they do not contribute to energy leakage. As a result, the non-Hermiticity of matrix $\boldsymbol{C}$ only originates from $\boldsymbol{C}_{rad}$ and $\boldsymbol{C}_{TE\text{-}TM}$.

At the 2nd-order $\Gamma$ point ($\boldsymbol{k}_{\parallel} =0$), we have $(m_x^2+n_y^2)=(m^2+n^2)=1$, and hence, the basic waves share identical vertical profile $\Theta_0(z)$ and wavenumber $k_0$. 
Matrix $\boldsymbol{K}$ turns into a constant $k_0^2$ and Eq.(\ref{eq:9}) becomes:
\begin{equation}
\label{eq:13}
(k^2-k_0^2)\mathbf{V}=\boldsymbol{C}\mathbf{V}
\end{equation}

As proved in Appendix \ref{App:B}, due to the $C_{4v}$ symmetry owned by the PC slab, the matrix $\boldsymbol{C}$ has a symmetric form as:
\begin{eqnarray}
\label{eq:14}
\boldsymbol{C}&=&\boldsymbol{A}+i\boldsymbol{B}\nonumber\\
&=&\left(
\begin{array}{cccc}
a &b &-c &c\\
b &a &c &-c\\
-c &c &a &b\\
c &-c &b &a
\end{array}
\right)+i\left(
\begin{array}{cccc}
d &d &0 &0\\
d &d &0 &0\\
0 &0 &d &d\\
0 &0 &d &d
\end{array}
\right)
\end{eqnarray}
where the matrix elements $a,b,c,d$ rely on a series of structural parameters, such as the filling factor, slab thickness, and the permittivity of the PC layer and upper/lower claddings. The CWT gives the explicit expressions of such matrix elements.

\section{\label{sec:level3} Mode degeneracy and band structure}
\subsection{\label{subsec:3.1} Condition of mode degeneracy at the $\Gamma$ point}

From the perspective of CWT, the accidental degeneracy is equivalent to the degeneracy of eigenvectors of matrix $\boldsymbol{C}$. The non-Hermiticity of the PC system is described by matrix $\boldsymbol{B}$. In order to deal with the radiative waves, we should consider the eigenvalues of matrix $\boldsymbol{C}$ rather than its Hermitian part matrix $\boldsymbol{A}$. 
 
\begin{figure}[!htp]
\centering
\includegraphics[width=8.5cm]{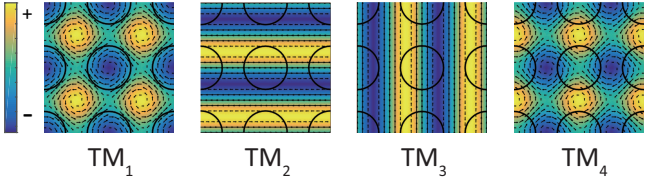}
\caption{Calculated field patterns of four band-edge modes (TM$_1$-TM$_4$) at the 2nd-order $\Gamma$ point. The vectorial arrows represent the magnetic components ($H_x$ and $H_y$) and  colors represent the electric component ($E_z$). The circles indicate the pillars on substrate.}
\label{fig:2}
\end{figure}

Solving the eigenvalue problem Eq.(\ref{eq:13}) leads to four band-edge modes TM$_{1\sim4}$ at the $\Gamma$ point, whose eigenstates are presented in Eq.(\ref{eq:15}). Their field patterns are illustrated in Fig. \ref{fig:2}.
\begin{eqnarray}
\label{eq:15}
\text{TM}_1:~~~\boldsymbol{C}\phi_1&=&(a-b+2c)\phi_1\nonumber\\
\text{TM}_2:~~~\boldsymbol{C}\phi_2&=&(a+b+i2d)\phi_2\\
\text{TM}_3:~~~\boldsymbol{C}\phi_3&=&(a+b+i2d)\phi_3\nonumber\\
\text{TM}_4:~~~\boldsymbol{C}\phi_4&=&(a-b-2c)\phi_4\nonumber
\end{eqnarray}
where
\begin{eqnarray}
\phi_{1}&=&\frac{1}{2}\left(\begin{array}{c}
-1\\1\\1\\-1 
\end{array}\right)~~~~~~~
\phi_{2}=\frac{1}{\sqrt{2}}\left(\begin{array}{c}
0\\0\\1\\1
\end{array}\right)\nonumber\\
\phi_{3}&=&\frac{1}{\sqrt{2}}\left(\begin{array}{c}
1\\1\\0\\0
\end{array}\right)~~~~~~~
\phi_{4}=\frac{1}{2}\left(\begin{array}{c}
1\\-1\\1\\-1
\end{array}
\right).\nonumber
\end{eqnarray}

Obviously, $\phi_1$ and $\phi_4$ are two degenerate eigenstates of matrix $\boldsymbol{B}$ with eigenvalue $0$, corresponding to real frequencies with infinite Q factor. On the contrary, $\phi_2$ and $\phi_3$ correspond to a non-zero eigenvalue of $2d$, leading to complex mode frequencies, namely, finite lifetime.

According to Eq.(\ref{eq:15}), the condition of mode degeneracy becomes quite straightforward. Since the two leaky modes TM$_2$ and TM$_3$ are degenerate by nature owing to their in-plane symmetry, accidental triple degeneracy can be realized by tuning the structural parameters to make them degenerate with a single high-Q mode, namely, mode TM$_1$. From the perspective of CWT, the real parts of their eigenvalues should be equal to each other, which gives the condition of mode degeneracy at the 2nd-order $\Gamma$ point:
\begin{eqnarray}
\label{eq:19}
&&a-b+2c=a+b\nonumber\\
&& \Longrightarrow ~~~~~~b=c
\end{eqnarray}

The detailed expressions of $b$ and $c$ can be found in Appendix \ref{App:B}. For a low-index-contrast structure, $b$ and $c$ can be simplified by only keeping the strongest coupling term $\boldsymbol{C}_{1D}$ but neglecting  $\boldsymbol{C}_{2D}$, $\boldsymbol{C}_{rad}$, $\boldsymbol{C}_{TE\text{-}TM}$, as:
\begin{eqnarray}
\label{eq:20}
&b_{1D}=\int_{PC}\kappa_{2,0}[-\frac{\partial^2}{\partial z^2}+\delta_A\frac{\partial}{\partial z}-\beta_0^2]\Theta_0(z)\Theta^*_0(z)dz\nonumber\\
&c_{1D}=\kappa_{1,-1}\beta_0^2\int_{PC}\Theta_0(z)\Theta^*_0(z)dz
\end{eqnarray}
where $\kappa_{m,n}=2f(1/\varepsilon_1-1/\varepsilon_q)J_1(\zeta)/\zeta$ with $\zeta=2\sqrt{f\pi(m^2+n^2)}$.  Apparently, $b$ and $c$ not only depend on the coupling coefficients $\kappa_{mn}\{f, \varepsilon_1,\varepsilon_q\}$, but also rely on the vertical profile $\Theta(z)\{h,f,\varepsilon_1,\varepsilon_2,\varepsilon_q\}$. For infinitely thick PCs stated in other works, the accidental degeneracy is realized by continuously varying the radius of the hole or pillar. However, the realistic PC slabs actually provide extra freedom to achieve $b=c$.

As presented in Fig. \ref{fig:1}(b), the element $b_{1D}$ depicts the feedback between basic waves $R_x$ and $S_x$ (also $R_y$ and $S_y$), whereas the element $c_{1D}$ presents the coupling between the basic waves propagating in orthogonal directions, for instance, $R_x$ and $R_y$. For the TE modes, the polarization of basic waves in $x$ and $y$ directions are normal to each other, which forbids the direct couplings between them. However, for the TM modes, the electric fields of the basic waves are always perpendicular to the transversal PC plane, and hence, can be coupled directly regardless of the propagating directions, leading to a non-zero value of $c$. As a result, the condition $b=c$ can be more easily fulfilled for the TM modes.

\subsection{\label{subsec:3.2}Linearity and isotropy of the  Dirac-cone like band near the $\Gamma$ point}

From the explicit expression of the triple degeneracy,we can analytically investigate the linearity and isotropy of the band near the 2nd-order $\Gamma$ point. Here we focus on a realistic structure shown in Fig. \ref{fig:1} (a) with structural parameters listed in Table I. The parameter set $(f,\varepsilon_{q})$ is tuned to realize this accidental triple degeneracy.
\renewcommand\arraystretch{1.2}
\begin{table}[htb]
\label{tab:2}
\centering
\caption{Structural parameters used in Fig.\ref{fig:3}}
\begin{tabular}{ccc}
\hline
Layer&Thickness& $\varepsilon(\mathbf{r})$ \\
\hline
upper-clad (liquid)&$\infty$&$\varepsilon_{q}$\\
PC (Si3N4/liquid)&$0.6a$&$ 4.0804/\varepsilon_q$\\
lower-clad (SiO2)&$\infty$&$2.1316$\\
\hline
\end{tabular}
\end{table}
\begin{figure}[!htp]
\centering
\includegraphics[width=8.5cm]{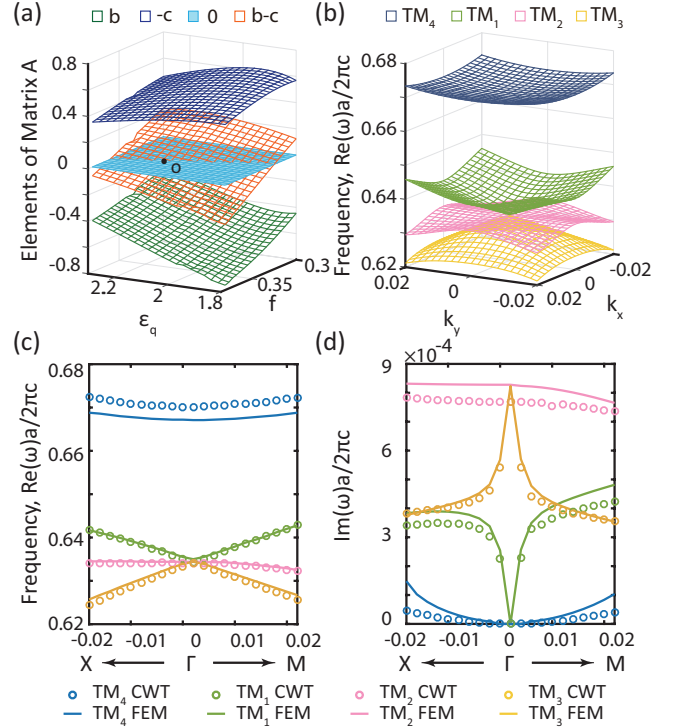}
\caption{(a) Elements of coupling matrix $\boldsymbol{C}$ in parameter space $\{f,\varepsilon_{q}\}$. Point $O$ is a specific point that we choose for the eigenvalue calculations. (b) Three-dimensional plot of the band structure calculated by CWT with structural parameters shown in Table I. The linear bands (green and orange) form a Dirac-like cone. The real (c) and imaginary (d) parts of the eigenvalues calculated by CWT (circles) and FEM (solid lines). The in-plane wavevector $\boldsymbol{k}_{\parallel}$ is along $\Gamma\text{-}X$ and $\Gamma\text{-}M$ directions.}
\label{fig:3}
\end{figure}

Fig. \ref{fig:3}(a) illustrates the elements $b$ and $c$ in the parameter space $\{f,\varepsilon_{q}\}$. The curvature $(b-c)$ intersects with the zero plane, which gives the line $b-c=0$, on which every point represents a set of $(f,\varepsilon_{q})$ for realizing triple degeneracy. Here we focus on the point $O (f=0.35,\varepsilon_{q}=2.15)$. Fig. \ref{fig:3}(b) presents the band structure of point $O$ in the $\boldsymbol{k}_{\parallel}$-space, and the details in $X\text{-}\Gamma\text{-}M$ direction are illustrated in Fig. \ref{fig:3}(c) and (d). The CWT results agree well with the finite-element method (FEM, COMSOL Multiphysics) simulations. The band structure clearly shows an accidental triple degeneracy at the 2nd-order $\Gamma$ point, which is formed by the degenerate leaky modes TM$_2$, TM$_3$ and high-Q mode TM$_1$. 


We investigate the band structure by assuming a small off-$\Gamma$ wave vector $\boldsymbol{k}_{\parallel}$. For a low-index-contrast Si$_3$N$_4$ structure, the coupling matrix $\boldsymbol{C}$ is almost invariant around the 2nd-order $\Gamma$ point, and hence, matrix $\boldsymbol{K}$ dominates the band gap at off-$\Gamma$ points.

It's easy to prove that (see Appendix \ref{App:C}):
\begin{equation}
\label{eq:21}
k_{0,mn}=k_0-\frac{\xi}{\varsigma}\frac{\beta_0}{k_0} (mk_x+nk_y)
\end{equation}
where $\xi$ and $\varsigma$ are the coefficients related to structural parameters. Therefore, the matrix $\boldsymbol{K}$ at an off-$\Gamma$ point is linear with respect to $k_x$ and $k_y$. Taking $\boldsymbol{K}$ as perturbation and applying the condition of triple degeneracy $b=c$ in matrix $\boldsymbol{C}$, the coupling matrix $\boldsymbol{D}$ under basis $\{\phi_3,\phi_4,\phi_1,\phi_2\}$ becomes:
\begin{eqnarray}
\label{eq:24}
\boldsymbol{D}=\left(
\begin{array}{cccc}
a+b+2id&\frac{1}{\sqrt{2}}\mu k_x &-\frac{1}{\sqrt{2}}\mu k_x &0\\
\frac{1}{\sqrt{2}}\mu k_x &a-3b &0 &\frac{1}{\sqrt{2}}\mu k_y\\
-\frac{1}{\sqrt{2}}\mu k_x &0 &a+b &\frac{1}{\sqrt{2}}\mu k_y\\
0 &\frac{1}{\sqrt{2}}\mu k_y &\frac{1}{\sqrt{2}}\mu k_y &a+b+2id
\end{array}
\right)\nonumber\\
\end{eqnarray}
where $\mu=-2 \xi\beta_0/\varsigma$.

The matrix above clearly depicts the couplingss brought by the off-$\Gamma$ wave vector $\boldsymbol{k}_{\parallel}$. As modes TM$_{1,2,3}$ are triply-degenerate and TM$_4$ is far above them, we reduce the matrix by neglecting the coupling terms with mode TM$_4$ as:
\begin{eqnarray}
\label{eq:25}
\boldsymbol{D}_{r}=\left(
\begin{array}{ccc}
a+b+2id&-\frac{1}{\sqrt{2}}\mu k_x &0\\
-\frac{1}{\sqrt{2}}\mu k_x &a+b &\frac{1}{\sqrt{2}}\mu k_y\\
0 &\frac{1}{\sqrt{2}}\mu k_y &a+b+2id
\end{array}
\right)
\end{eqnarray}

One eigenvalue of $\boldsymbol{D}_r$ is $\lambda_2=a+b+i2d$, corresponding to the flat band whose real part remains $a+b$ and imaginary part remains $2d$. Despite this trivial solution, the other two eigenvalues are
\begin{eqnarray}
\label{eq:27}
\lambda_{1,3}=(a+b+id)\pm\sqrt{\frac{\mu^2}{2}(k_x^2+k_y^2)-d^2}
\end{eqnarray}
Clearly $\lambda_{1,3}$ only depend on $|\boldsymbol{k}_{\parallel}|=\sqrt{k_x^2+k_y^2}$ regardless of the specific direction, which means the mode frequencies are isotropic in arbitrary direction. 

Eq.(\ref{eq:27}) indicates that, there exists an off-$\Gamma$ ring satisfying $\sqrt{k_x^2+k_y^2}=\sqrt{2}d/\mu$ on which $\lambda_1$ and $\lambda_3$ are fully degenerate. These off-$\Gamma$ degenerate points are EPs that we will elaborate in the following sections. It is noteworthy that, as long as the radiation exists, the ring of EPs would always exist around the 2nd-order $\Gamma$ point, making the band discontinuous and not cone-like. In other words, there's no rigorous cone dispersion in non-Hermitian photonic crystal slab with $C_{4v}$ symmetry.

However, if $d$ is sufficiently small, the EPs ring would be very close to 2nd-order $\Gamma$ point, and the overall band would still look like a cone. For instance, the band illustrated in Fig. \ref{fig:3}(c) clearly exhibits a cone-like dispersion. In this case, back to Eq.(\ref{eq:27}), $d$ could be neglected and $\lambda_{1,3}$ are both linear with the same slope coefficient of  $\mu/\sqrt{2}$, which indicates that the top and bottom halves of the cone share the same linearity. 

Some interesting details can be noticed from Fig. \ref{fig:3}(c). For example, the linearity of the band in $X\text{-}\Gamma$ direction remarkably degrades when deviating from 2nd-order $\Gamma$ point. Moreover, in $\Gamma\text{-}M$ direction, the leaky mode TM$_2$ no longer remains flat but behaves as ripple-like dispersion which could easily be observed in Fig. \ref{fig:3}(b). We believe this band distortion is due to the influence of mode TM$_4$. Such mode has been neglected in previous studies of Hermitian systems, and also been dropped for simplicity in our reduced matrix $\boldsymbol{D}_r$. The couplings brought by TM$_4$ are proportional to $k_x$ or $k_y$, and hence, significant band distortion will occur at large off-$\Gamma$ wave vectors. On the other hand, for a high-index-contrast structure, or strong optical confinement, the linear region of the cone-like band will be relatively small since the band could be twisted by the strong coupling strength.

\subsection{\label{subsec:3.3} Ring of exceptional points}
As mentioned, the EPs ring where $\lambda_1$ and $\lambda_3$ are fully degenerate is inevitable in non-Hermitian PCs. As stated above, Eq.(\ref{eq:27}) indicates the EPs ring appearing at the off-$\Gamma$ wave vectors satisfying:
\begin{equation}
\label{eq:30}
|\boldsymbol{k}_{\parallel}|=\frac{\sqrt{2} d}{\mu}
\end{equation}
Obviously, there is an infinite number of EPs on the ring which separates the whole plane into two sections. Inside this ring, the real frequencies of $\lambda_1$ and $\lambda_3$ are degenerate as $a+b$, while outside this ring, the imaginary parts are degenerate as a value of $d$. The ring is a collection of singular points and its size depends on the strength of radiation loss $d$ and coefficient $\mu$.

\begin{figure}[!htp]
\centering
\includegraphics[width=8.5cm]{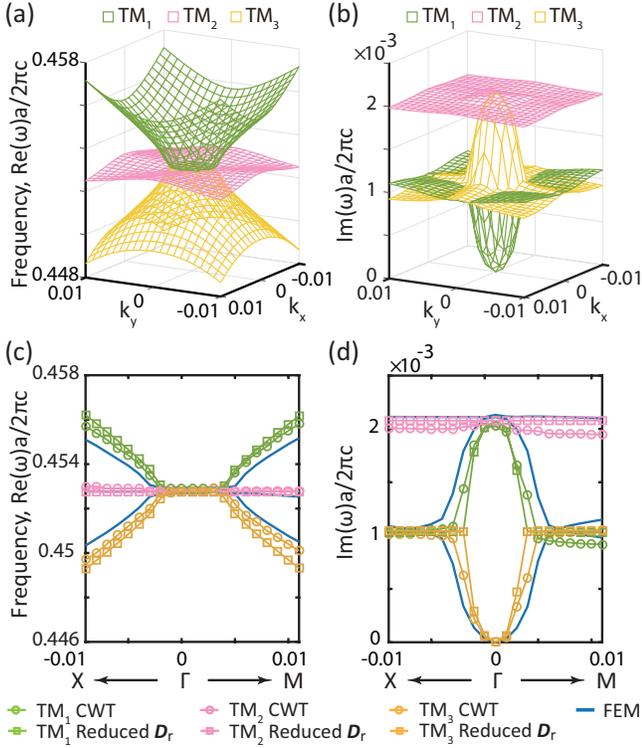}
\caption{Three-dimensional plot of the real (a) and imaginary (b) parts of the eigenvalues calculated by CWT with structural parameters shown in Table II. The real (c) and imaginary (d) parts of the eigenvalues with $\boldsymbol{k}_{\parallel}$ along the specific $\Gamma\text{-}X$ and $\Gamma\text{-}M$ directions, calculated by CWT (circles), CWT with reduced coupling matrix $\boldsymbol{D}_r$ (squares), and FEM (solid lines).}
\label{fig:4}
\end{figure}

Here we present a realistic structure with relatively strong radiation loss (Q$_v\approx 110$), in order to demonstrate a distinct and clear EPs ring. The structural parameters are listed in Table II.
\renewcommand\arraystretch{1.2}
\begin{table}[htb]
\label{tab:3}
\centering
\caption{Structural parameters used in Fig.\ref{fig:4}}
\begin{tabular}{cccc}
\hline
Layer&Thickness& $\varepsilon(\mathbf{r})$&$f$ \\
\hline
upper-clad (liquid)&$\infty$&$4.00$&-\\
PC (Si/liquid)&$0.4a$&$ 12.11/4.00$&$0.28$\\
lower-clad (Si3N4)&$\infty$&$4.0804$&-\\
\hline
\end{tabular}
\end{table}

Fig. \ref{fig:4} (a) and (b) illustrate the complex band structure in $\boldsymbol{k}_{\parallel}$-space. The EPs ring can be clear observed from both the real and imaginary parts of the band. The details of the band in  $X\text{-}\Gamma\text{-}M$ direction are presented in Fig. \ref{fig:4} (c) and (d). The CWT results are in good agreement with FEM simulations, which confirms the validity of the CWT model.

Eq.(\ref{eq:27}) is derived from the reduced coupling matrix $\boldsymbol{D}_r$ neglecting the couplings with mode TM$_4$. As a result, the EPs ring is exactly a circle, in other words, isotropic in the $\boldsymbol{k}_{\parallel}$-space. From Fig. \ref{fig:4} (c) and (d), we find the results given by $\boldsymbol{D}_r$ agree well with the CWT results and FEM results. However, similar to the distortion of band in Fig. \ref{fig:3}, the couplings between TM$_4$ and $\text{TM}_{1,2,3}$ also degrade the isotropy of the ring.

\section{\label{sec:level4} Topological properties of the band with triple degeneracy}

In this section, we investigate the topological properties of band with triple degeneracy from reduced coupling matrix $\boldsymbol{D}_r$ in several specific parameter spaces.
\subsection{\label{subsubsec:4.1} Zero Berry Phase of band structure in $\boldsymbol{k}_{\parallel}$-space}

\begin{figure}[!htp]
\centering
\includegraphics[width=8.5cm]{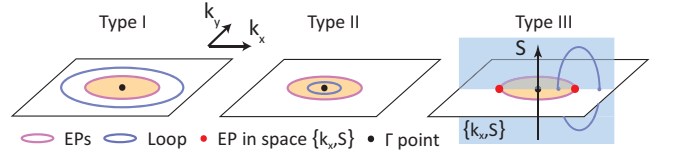}
\caption{Three different loops. Type \uppercase\expandafter{\romannumeral1} (outside the EPs ring) and Type \uppercase\expandafter{\romannumeral2} (inside the EPs ring) loops are in the $\boldsymbol{k}_{\parallel}$-space resulting in no topological interchange and zero Berry phase; Type \uppercase\expandafter{\romannumeral3} loop is in parameter space $\{k_x,S\}$ where $S$ refers to a structural parameter.}
\label{fig:5}
\end{figure}

As we stated above, in non-Hermitian case, there always exists a EPs ring around the 2nd-order $\Gamma$ point. On the ring, the coupling matrix becomes defective. Due to the position with respect to the EPs ring, there are three typical different loops in $\boldsymbol{k}_{\parallel}$-space shown in Fig. \ref{fig:5}.

First we consider type \uppercase\expandafter{\romannumeral1} and type \uppercase\expandafter{\romannumeral2} loop. It's clear that there are no singular points on these loops. We revisit the Berry phase that may be accumulated along these loops, since it is an important indicator to distinguish whether the cone-like band follows the Dirac equation. The Berry phase in $\boldsymbol{k}_{\parallel}$-space for a non-Hermitian system is defined as \cite{32}
\begin{equation}
\label{eq:31}
\gamma=i\oint\left\langle\Phi(\boldsymbol{k}_{\parallel})|\nabla_{\boldsymbol{k}_{\parallel}}|\Psi(\boldsymbol{k}_{\parallel})\right\rangle d\boldsymbol{k}_{\parallel}+\eta
\end{equation}
Where $e^{i\eta(T)}=\left\langle\Phi(\boldsymbol{k}_{\parallel}(0))|\Psi(\boldsymbol{k}_{\parallel}(T))\right\rangle$ and $T$ refers to the end of loop. The left vector $\Phi$ is the eigenvector in the dual space. From the perspective of CWT, the explicit wave functions of the degenerate modes can be chosen as the eigenvectors of coupling matrix. Here we calculate the $\Psi$ from reduced coupling matrix $\boldsymbol{D}_r$ to simplify the evaluation. This is valid because we only concern a small region around the 2nd-order $\Gamma$ point where the coupling terms with mode TM$_4$ is negligible.

Here we take $|\Psi_1\rangle$ with eigenvalue $\lambda_1$ in Eq.(\ref{eq:27}) as an example, whose left vector $|\Phi_1\rangle$ satisfies $\lambda_1^*|\Phi_1\rangle=\boldsymbol{D}_r^\dag|\Phi_1\rangle$. Obviously $\boldsymbol{D}_r$ is a symmetric non-Hermitian matrix, so we have $\boldsymbol{D}_r^\dag=\boldsymbol{D}_r^*$ and the left vector can be chosen as $|\Phi_1\rangle=|\Psi_1^*\rangle$. Differentiation of the orthogonalization relation $\left\langle\Psi_1^*(\boldsymbol{k}_{\parallel})|\Psi_1(\boldsymbol{k}_{\parallel})\right\rangle=1$ yields
\begin{eqnarray}
\label{eq:33}
& &\nabla_{\boldsymbol{k}_p}\left\langle\Psi_1^*(\boldsymbol{k}_{\parallel})|\Psi_1 (\boldsymbol{k}_{\parallel})\right\rangle\nonumber\\
&=&\left\langle\nabla_{\boldsymbol{k}_{\parallel}}\Psi_1^*(\boldsymbol{k}_{\parallel})|\Psi_1 (\boldsymbol{k}_{\parallel})\right\rangle+\left\langle\Psi_1^*(\boldsymbol{k}_{\parallel})|\nabla_{\boldsymbol{k}_{\parallel}}\Psi_1 (\boldsymbol{k}_{\parallel})\right\rangle\nonumber\\
&=&2\left\langle\Psi_1^*(\boldsymbol{k}_{\parallel})|\nabla_{\boldsymbol{k}_{\parallel}}\Psi_1 (\boldsymbol{k}_{\parallel})\right\rangle=0.
\end{eqnarray}
So the Berry connection $\left\langle\Phi_1(\boldsymbol{k}_{\parallel})|\nabla_{\boldsymbol{k}_{\parallel}}|\Psi_1(\boldsymbol{k}_{\parallel})\right\rangle=0$, and the Berry phase is solely dominated by $\eta$ in Eq.(\ref{eq:31}). Clearly, there's no modal interchange along the Type \uppercase\expandafter{\romannumeral1} and Type \uppercase\expandafter{\romannumeral2} loops and no phase change of the instantaneous eigenvector $\Psi_1(\boldsymbol{k}_{\parallel})$. So phase $\eta$ vanishes and Berry phase remains zero. For eigenvector $|\Psi_3\rangle$ with eigenvalue $\lambda_3$, the result is the same.

Zero Berry phase indicates that, encircling around the $\Gamma$ point in the $\boldsymbol{k}_{\parallel}$-space dose not accumulate any extra geometric phase, no matter the loop is inside the EP ring or outside the EP ring. Therefore, the band structure at the $\Gamma$ point under $C_{4v}$ symmetry can not be described by the Dirac equation. In other words, the cone-like dispersion we observe at 2nd-order $\Gamma$ point with $C_{4v}$ symmetry is actually not a Dirac cone but Dirac-\textit{like} cone. Compared to Dirac cone in Graphene, it does not give rise to some unique properties like anti-localization when disorder exists. Nevertheless, the Dirac-like cone still 
possesses good linearity and isotropy near the 2nd-order $\Gamma$ point with small radiation.

\subsection{\label{subsec:4.2} Modal interchange and non-zero Berry phase in parameter space $\{k_x,S\}$}

Though Berry phase remains zero in $\boldsymbol{k}_{\parallel}$-space, non-zero Berry phase may be realized in other parameter spaces. We consider a loop in space $\{k_{\parallel},S\}$ where $S$ refers to any structural parameter and $k_{\parallel}$ refers to a specific direction in $\boldsymbol{k}_{\parallel}$-space since the band structure is isotropic in the vicinity of 2nd-order $\Gamma$ point. Here we take $\{k_y=0, S=f\}$ as an example. When $f$ varies, the condition of triple degeneracy $b=c$ no longer holds and we denote $S=b-c$, which is equivalent to parameter $f$. 

Space $\{k_x,S\}$ is shown in Fig. \ref{fig:5} on which loop is denoted as Type \uppercase\expandafter{\romannumeral3}. Clearly we find space $\{k_x,S\}$ contains two EPs at $(\pm\sqrt{2}d/\mu,0)$.

When $f$ varies, the reduced coupling matrix becomes 
\begin{eqnarray}
\label{eq:36}
\boldsymbol{D}_r^f=\left(
\begin{array}{ccc}
a+b+2id&-\frac{1}{\sqrt{2}}\mu k_x &0\\
-\frac{1}{\sqrt{2}}\mu k_x &a-b+2c &\frac{1}{\sqrt{2}}\mu k_y\\
0 &\frac{1}{\sqrt{2}}\mu k_y &a+b+2id
\end{array}
\right)
\end{eqnarray}
the non-trivial eigenvalues of $\boldsymbol{D}_r^f$ with $k_y=0$ are
\begin{equation}
\label{eq:37}
\lambda^f_{1,3}=(a+c+id)\pm\sqrt{\frac{\mu^2}{2}k_x^2+(S+id)^2}
\end{equation}

\begin{figure}[!htp]
\centering
\includegraphics[width=8.5cm]{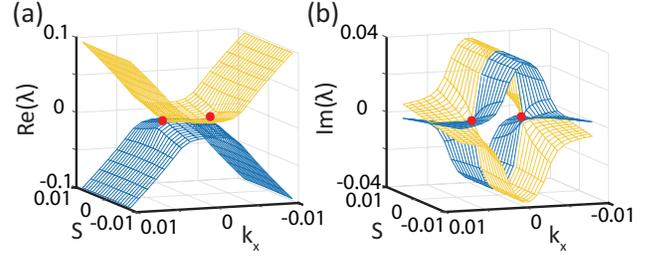}
\caption{The real (a) and imaginary parts (b) of the eigenvalues $\lambda_{1,3}$. The red dots denote two EPs at $(\pm \sqrt{2}d/\mu,0)$. }
\label{fig:6}
\end{figure}

Fig. \ref{fig:6} gives the Riemann surfaces of $\lambda^f_{1,3}$ in parameters space $\{k_x,S\}$, from which two EPs at $(\pm \sqrt{2}d/\mu,0)$ can also be clearly observed. In fact, the two EPs are the branch points at which the two sheets of Riemann surfaces intersect. On the line $S=0$, the real parts are degenerate for $|k_x|<\sqrt{2}d/\mu$, and the imaginary parts are degenerate for $|k_x|>\sqrt{2}d/\mu$. 

\begin{figure*}[!htp]
\centering
\includegraphics[width=17.5cm]{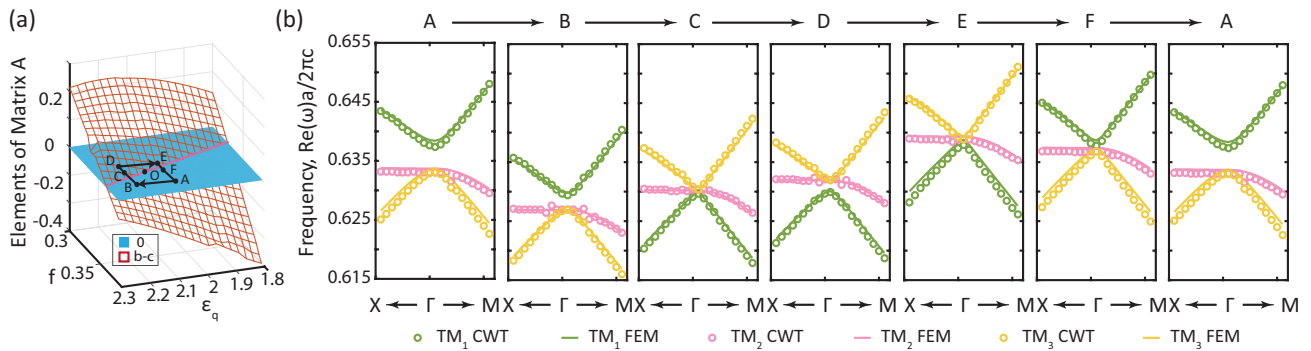}
\caption{(a). The parameter space $\{f, \varepsilon_{q}\}$ and the loop $\{A-B-C-D-E-F-A\}$ around the degenerate point $O$; (b). band structure of points on the loop. The parameter sets $\{f,\varepsilon_q\}$ for A, B, C, D, E and F points are $\{0.37, 2.10\}$,  $\{0.37,2.22\}$, $\{0.35, 2.22\}$,  $\{0.34, 2.22\}$, $\{0.34, 2.10\}$, and $\{0.35, 2.10\}$, respectively, with other parameters listed in Table I.}
\label{fig:7}
\end{figure*}

The loop evolving in the parameter space $\{k_x,S\}$ can be topologically different according to how the EPs are encircled. When no EP is encircled, the eigenvalues simply return to their initial values. When a single EP is encircled, both the real and imaginary parts of the eigenvalues cross to the other sheet after finishing the loop, as a result, the eigenvalues interchange with each other. After two cycles the eigenvalues return to theirselves. Further, when both two EPs are encircled, the real parts of eigenvalues stay on the same sheet while the imaginary parts cross to the other sheet but will then cross back. After finishing the loop, the eigenvalues return to their initial values.

The eigenvectors corresponding to $\lambda_{1,3}^f$ of $\boldsymbol{D}_r^f$ are
\begin{eqnarray}
\label{eq:38}
u_1=\frac{1}{N_-}\left(\begin{array}{c}
\frac{1}{\sqrt{2}}\mu k_x\\S+id-M\\0
\end{array}\right)u_3=\frac{1}{N_+}\left(\begin{array}{c}
\frac{1}{\sqrt{2}}\mu k_x\\S+id+M\\0
\end{array}\right)\nonumber\\
\end{eqnarray}
where $N_{\pm}=\sqrt{\mu^2k_x^2/2+(S+id\pm M)}=\sqrt{2M^2\pm 2M(S+id)}$ and $M=\sqrt{\mu^2k_x^2/2+(S+id)^2}$. Since coupling matrix $\boldsymbol{D}_r^f$ is symmetric, as we stated in last section, the left vectors could be chosen as $|v_{1,3}\rangle=|u^*_{1,3}\rangle$. In order to evaluate the Berry phase, we analyze the behavior of the eigenvectors $u_1, u_3$ in space $\{k_x,S\}$ by assuming a small circle of radius $r$ around a single EP in Fig. \ref{fig:6} \cite{34}, chosen here as $(\mu k_x/\sqrt{2}=+d,S=0)$:
\begin{equation}
\label{eq:39}
\frac{\mu k_x}{\sqrt{2}}=+d+r\cos\varphi ~~~~~~~ S=r\sin\varphi
\end{equation}
With the approximation $r\ll d$, the two eigenvectors $u_{1,3}$ could be represented in the form:
\begin{eqnarray}
\label{eq:39}
u_1=\left(\begin{array}{c}
-\sin\theta\\\cos\theta\\0
\end{array}\right) ~~~~~~~ u_3=\left(\begin{array}{c}
\cos\theta\\\sin\theta\\0
\end{array}\right)
\end{eqnarray}
where 
\begin{eqnarray}
\label{eq:40}
\tan\theta&=&\frac{id+r\sin\varphi+\sqrt{r^2+2dre^{i\varphi}}}{d+r\cos\varphi}\nonumber\\
&=&i+\sqrt{\frac{2r}{d}}e^{i\frac{\varphi}{2}}+O(r)
\end{eqnarray}
and further we have
\begin{eqnarray}
\label{eq:41}
Re(\theta)&=&Re(\arctan(i+\sqrt{\frac{2r}{d}}e^{\frac{\varphi}{2}}))=Re(\arctan(z))\nonumber\\
&=&Re(-\frac{i}{2}\ln\frac{1+iz}{1-iz})=\frac{1}{2}(\frac{\pi}{2}+\frac{\varphi}{2})
\end{eqnarray}
Eq.(\ref{eq:41}) indicates the behavior of eigenvectors:
\begin{equation}
\label{eq:42}
u_1\stackrel{2\pi}{\longrightarrow}-u_3\stackrel{2\pi}{\longrightarrow}-u_1 ~~~~~~~
u_3\stackrel{2\pi}{\longrightarrow}+u_1\stackrel{2\pi}{\longrightarrow}-u_3
\end{equation}
Obviously, the eigenvectors are interchanged after a single circle encircling one EP and one of them gain extra phase $\pi$. After second cycle both eigenvectors are reconstructed with extra phase $\pi$. Further, we can infer that for a single loop encircling both two EPs, both eigenvectors accumulate a extra phase $\pi$ too. As the coupling matrix $\boldsymbol{D}_r^f$ is symmetric and Berry phase is dominated by phase $\eta$ in Eq.(\ref{eq:31}), the geometric phase is 
\begin{equation}
\label{eq:43}
\gamma_{1,3}=\pi
\end{equation}
both for two circle encircling a single EP and one circle encircling both EPs. 

What we should point out is that the approximation $r\ll d$ we take in considerations above is only used to investigate the behavior of eigenvectors in a simple way. Virtually because of the continuity of the band structure except EPs in Fig.\ref{fig:6}, the conclusion Eq.(\ref{eq:43}) are exact regardless of the specific loop.

\subsection{\label{subsec:4.3} Modal exchange in parameter space $\{f,\varepsilon_q\}$}

Further, we consider a structural parameter space, i.e space $\{f,\varepsilon_q\}$ shown in Fig. \ref{fig:7} (a). In this parameter space exists line $b=c$ corresponding to condition of triple degeneracy. It is noteworthy that, because of the uncertainty in fabricating realistic devices, the working points in this space would always slightly deviate from the exact degenerate condition $b=c$. Due to the uncertainty of the measurement, it is also difficult to distinguish whether the rigorous degenerate condition $b=c$ has been met.  

When the parameters slightly deviate from $b=c$, the sign of $b-c$ determines which mode (mode TM$_1$ or TM$_3$) lying on the top half of the cone. More specifically, for the points in parameter space with $b>c$, mode TM$_1$ forms the bottom half; while for the points with $b<c$, mode TM$_3$ turns to be the bottom half. 

Such phenomena is illustrated in Fig. \ref{fig:7} (b). When system evolves adiabatically along a loop encircling a rigorous degenerate point $O$, the system would cross the $b-c$ plane twice and return to its initial state. During this loop encircling, we can observe that the two bands forming the up/bottom halves of the cone switch twice, and eventually come back to the initial state. We refer to this evolution as a topological exchange of the modes.

\section{\label{sec:level5}CONCLUSION}

In this work, we present an analytical study of the mode degeneracy with TM-like polarization in a non-Hermitian photonic crystal slab with $C_{4v}$ symmetry. The CWT framework provides an analytical perspective to comprehensively understand the physics of a non-Hermitian photonic crystal system. We introduce the TE-TM couplings to improve the accuracy and extend CWT framework to depict the analytical wave functions and complex band structure through reduced coupling matrix.

From the CWT framework, the elements of the coupling matrix are evaluated , and hence, the condition of realizing accidental triple degeneracy is derived. Comparing to the Hermitian PCs that extend infinitely along $z$ direction, the realistic non-Hermitian PCs provide extra freedoms in tuning the structural and material parameters to fulfill the accidental degeneracy condition. The ring of EPs appears around 2nd-order $\Gamma$ point inevitably with triple degeneracy in non-Hermitian system. The band structure exhibits a cone-like dispersion with weak radiation loss. When the radiation loss become strong, the ring of EPs becomes distinct and easy to observe.

Furthermore, the linearity and isotropy, as well as the topological properties on the triple degeneracy, have been studied with the reduced coupling matrix. The analytical study reveals that, the cone-like band in a non-Hermitian PC with weak radiation loss possesses good linearity and isotropy in the vicinity of the $\Gamma$ point, which is similar to a real Dirac cone. The ring of EPs with strong radiation loss also owns good isotropy. 

Since the CWT gives a set of explicit wave functions of the degenerate modes, we can use them to calculate the Berry phase directly. The Berry phase remains zero at the center of the Brillouin zone, even the system is non-Hermitian. It indicates that the band structure induced by the accidental triple degeneracy with weak radiation loss is only Dirac-\textit{like} cone. Different from Graphene with real Dirac-cone, the PC with Dirac-like cone doesn't obey Dirac equation, and doesn't give rise to some unique properties like anti-localization against disorder, too. However, analytical study reveals that a loop in parameter space $\{k_x,S\}$ continuously and adiabatically encircling EPs gives rise to a modal interchange phenomenon and results in a non-zero Berry phase $\pi$, where S refers to a structural parameter like $f$. Moreover, a modal exchange of the band will occur when the parameters evolve as a loop encircling a rigorous degenerate point in the structural parameter space $\{f,\varepsilon_q\}$.

\section*{Acknowledgments}
This work was partly supported by the National Natural Science Foundation of China under Grant 61320106001, 61575002 and the State Key Laboratory of Advanced Optical Communication Systems and Networks, China.
Yong Liang is supported by the ETH Zurich Postdoctoral Fellowship Program (No. FEL-27 14-2) cofounded by the Marie Curie Actions for People COFUND Program.

\appendix\section{the boundary conditions} \label{App:A}
The representation in details of in-plane coupling items in Eq.(\ref{eq:2.a})-(\ref{eq:2.c}) are
\begin{eqnarray}
\label{eq:A.1}
\Pi_p&=&-\frac{\partial^2}{\partial z^2}(n_y H_{x,m' n'}-m_x H_{y,m' n'})+(m_x^2+n_y^2)\beta_0^2\nonumber\\
& &(n'_yH_{x,m' n'}-m'_xH_{y,m' n'})+il_1\beta_0\frac{\partial}{\partial z}H_{z, m' n'}\nonumber\\
\Xi_p&=&-\frac{\partial}{\partial z}(m_x H_{x,m' n'}+n_y H_{y,m' n'})-l_2\beta_0\frac{\partial}{\partial z}H_{z,m' n'}\nonumber\\
\Upsilon_p&=&l_2\beta_0^2H_{z,m' n'}-i\beta_0\frac{\partial}{\partial z}(m_x H_{x,m' n'}+n_y H_{y,m' n'})\nonumber\\
\end{eqnarray}

And the surface coupling items have the representation in details as:
\begin{eqnarray}
\label{eq:A.2}
\Pi_s&=&\delta_h\frac{\partial}{\partial z}(n_y H_{x,m' n'}-m_x H_{y,m' n'})-i\delta_hl_1\beta_0H_{z,m'n'}\nonumber\\
\Xi_s&=&\delta_h\frac{\partial}{\partial z}(m_x H_{x,m' n'}+n_y H_{y,m' n'})+i\delta_hl_2\beta_0H_{z,m'n'}\nonumber\\
\end{eqnarray}
where $\delta_h=\delta(z-h)-\delta(z+h)$, $l_1=m_x n'_y-n_y m'_x$ and $l_2=m_x m'_x+n_y n'_y$.

The boundary conditions can be derived from the integral on Eq.(\ref{eq:2.a}) and (\ref{eq:2.b}) over the upper and down boundaries $(h-\delta,h+\delta)$  and $(-h-\delta,-h+\delta)$ with $\delta\to 0$. Since the the Dirac function $\delta_h$ terms in $\Pi_s, \Xi_s$ has been treated as surface coupling,  the discontinuity only comes from the $\delta_K$ terms in operator $\mathcal{H}$ and $\mathcal{H}_{\delta}$. We have:

\begin{widetext}
\begin{eqnarray}
\label{eq:A.1}
& \textrm{for Eq.(\ref{eq:2.a})} :&\\
& &\kappa_b\frac{\partial}{\partial z}(n_y H_{x,m n}-m_x H_{y,m n})|_{h+0}=\kappa_a\frac{\partial}{\partial z}(n_y H_{x,m n}-m_x H_{y,m n})|_{h-0}\nonumber\\
& &\kappa_a\frac{\partial}{\partial z}(n_y H_{x,m n}-m_x H_{y,m n})|_{-h+0}=\kappa_c\frac{\partial}{\partial z}(n_y H_{x,m n}-m_x H_{y,m n})|_{-h-0}\nonumber
\label{eq:A.2}
\end{eqnarray}
\begin{eqnarray}
& \textrm{for Eq.(\ref{eq:2.b})} :& \\
& &[\kappa_b+(\kappa_b-\kappa_a)\frac{\kappa_b\beta_0^2(m_x^2+n_y^2)}{k^2-\kappa_b\beta_0^2(m_x^2+n_y^2)}]\frac{\partial}{\partial z}(m_x H_{x,m n}+n_y H_{y,m n})|_{h+0}=\kappa_a\frac{\partial}{\partial z}(m_x H_{x,m n}+n_y H_{y,m n})|_{h-0}\nonumber\\
& &[\kappa_c+(\kappa_c-\kappa_a)\frac{\kappa_c\beta_0^2(m_x^2+n_y^2)}{k^2-\kappa_c\beta_0^2(m_x^2+n_y^2)}]\frac{\partial}{\partial z}(m_x H_{x,m n}+n_y H_{y,m n})|_{-h-0}=\kappa_a\frac{\partial}{\partial z}(m_x H_{x,m n}+n_y H_{y,m n})|_{-h+0}\nonumber
\end{eqnarray}
\end{widetext}
where the terms of $(\kappa_b-\kappa_a)\{\kappa_b\beta_0^2(m_x^2+n_y^2)/(k^2-\kappa_b\beta_0^2(m_x^2+n_y^2))\}$ and $(\kappa_c-\kappa_a)\{\kappa_c\beta_0^2(m_x^2+n_y^2)/(k^2-\kappa_c\beta_0^2(m_x^2+n_y^2))\}$ come from the $\delta_K H_{z,m n}$ terms in $\mathcal{H}_\delta$ operator, which is a result of the TE-TM coupling. Because there is no Dirac function in $\mathcal{H}_z$ operator, the boundary conditions for Eq.(\ref{eq:2.c}) is continuous.

\begin{figure}[htp]
\centering
\includegraphics[width=8.5cm]{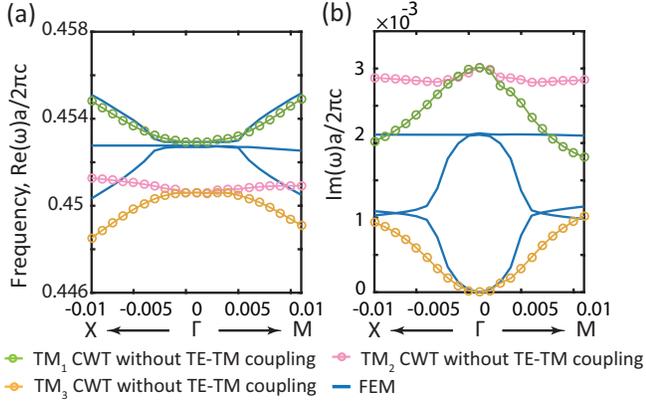}
\caption{The real (a) and imaginary (b) parts of the eigenvalues with $\boldsymbol{k}_{\parallel}$ along the specific $\Gamma\text{-}X$ and $\Gamma\text{-}M$ directions, calculated by CWT without TE-TM couplings (circles) and FEM (blue lines)}
\label{fig:A.1}
\end{figure}

Coupling Eqs.(\ref{eq:2.a})-(\ref{eq:2.c}) include TE-TM couplings introduced by component $H_{z,m n}$. To highlight the importance of such TE-TM couplings, we show in Fig. \ref{fig:A.1} the band structure evaluated by CWT without TE-TM couplings. The structural parameters are the same as those used for Fig. \ref{fig:4}, as listed in Table II. By comparing to Fig. \ref{fig:4} (which includes the TE-TM couplings), we find that the CWT results without TE-TM couplings cannot reproduce the FEM results and describe the feature of the EPs. Therefore, including the TE-TM couplings is critical for the accurate modeling of Dirac-like-cone dispersion within non-Hermitian PC systems.

\section{the matrix $\boldsymbol{C}$ at the $\Gamma$ point}\label{App:B}
Owing to the $C_{4v}$ symmetry at the $\Gamma$ point, the coupling matrix $\boldsymbol{C}$ should be symmetric to some extent. As stated above, the coupling matrix $\boldsymbol{C}$ can be divided into four parts $\{\boldsymbol{C}_{1D}, \boldsymbol{C}_{rad}, \boldsymbol{C}_{2D}, \boldsymbol{C}_{TE\text{-}TM}\}$. First part $\boldsymbol{C}_{1D}$ has a form as:
\begin{equation}
\label{eq:B.1}
\boldsymbol{C}_{1D}=\left(
\begin{array}{cccc}
0 &\chi_{2,0} &-\kappa_{1,-1}\beta_0^2 &\kappa_{1,1}\beta_0^2\\
\chi_{-2,0} &0 &\kappa_{-1,-1}\beta_0^2 &-\kappa_{-1,1}\beta_0^2\\
-\kappa_{-1,1}\beta_0^2 &\kappa_{1,1}\beta_0^2 &0 &\chi_{0,2}\\
\kappa_{-1,-1}\beta_0^2 &-\kappa_{1,-1}\beta_0^2 &\chi_{0,-2} &0
\end{array}
\right)\\
\end{equation}
\newline
where $\chi_{m,n}=\int_{PC}\kappa_{m,n}\{[-\partial^2/\partial z^2+\delta_A\partial/\partial z-\beta_0^2]\Theta_0(z)\}\Theta_0^*(z)dz$.

The second part $\boldsymbol{C}_{rad}$ corresponds to the couplings via the radiative waves, which can be evaluated with Green's function. Further, we can cacluate $\boldsymbol{C}_{rad}$ as 
\begin{equation}
\label{eq:B.2}
\boldsymbol{C}_{rad}=\left(
\begin{array}{cccc}
u_{10,-10} &u_{10,10} &0 &0\\
u_{-10,-10} &u_{-10,10} &0 &0\\
0 &0 &u_{01,0-1} &u_{01,01}\\
0 &0 &u_{0-1,0-1} &u_{0-1,01}
\end{array}
\right)
\end{equation}
where $u_{mn,pq}=\iint_{PC}\kappa_{m,n}\kappa_{p,q}\{[-\partial^2/\partial z^2+\delta_A\partial/\partial z]\cdot G_{0,-}(z,z')[(\partial^2/\partial z'^2-\delta'_A\partial/\partial z')\Theta_0(z')]\}\Theta_0^*(z)dz'dz$ and Green's function $G_{0,-}(z,z')$ is the solution of $\mathcal{H}(m_x=0,n_y=0)G_{0,-}(z,z')=-\delta(z,z')$, with applying the boundary conditions Eq.(\ref{eq:A.1}) in Appendix \ref{App:A}.

The third part $\boldsymbol{C}_{TE\text{-}TM}$ depicts the couplings between the basic waves $\{H_{x,1 0},H_{x,-1 0},H_{y,0 1}, H_{y,0 -1}\}$ and $\{H_{z,1 0},H_{z,-1 0},H_{z,0 1}, H_{z,0 -1}\}$. Similarly, we have: 
\begin{widetext}
\begin{equation}
\begin{aligned}
\label{eq:B.3}
&\boldsymbol{C}_{TE\text{-}TM}=\left(
\begin{array}{cccc}
v_{10,10}^{01}+v_{10,10}^{0-1} &u_{10,-10}^{01}+u_{10,-10}^{0-1} &0 &0\\
v_{-10,10}^{01}+v_{-10,10}^{0-1} &u_{-10,-10}^{01}+u_{-10,-10}^{0-1} &0 &0\\
0 &0 &v_{01,01}^{10}+v_{01,01}^{-10} &v_{01,0-1}^{10}+v_{01,0-1}^{-10}\\
0 &0 &v_{0-1,01}^{10}+v_{0-1,01}^{-10} &v_{0-1,0-1}^{10}+v_{0-1,0-1}^{-10}
\end{array}
\right)&\\
&\text{where}&\\
&v_{mn,pq}^{st}=\kappa_{\substack{m-s\\n-t}}\kappa_{\substack{s-p\\t-q}}\iint_{PC}\{[-\frac{\partial^2}{\partial z^2}+\beta_0^2+\delta_A\frac{\partial}{\partial z}]G_{st,+}(z,z')[(\frac{\partial^2}{\partial z'^2}-\delta'_A\frac{\partial}{\partial z'})\Theta_0(z')]+\delta_A\beta_0^2\frac{\partial}{\partial z'}\Theta_0(z')G_{st,0}\}\Theta_0^*(z)dzdz'&
\end{aligned}
\end{equation}
\end{widetext}
Green's function $G_{st,+}(z,z')$ is the solution of $\mathcal{H}(m_x=s+\Delta x,n_y=t+\Delta y)G_{st,+}(z,z')=-\delta(z,z')$ with applying the boundary condition Eq.(\ref{eq:A.2}).  $G_{st,0}(z,z')$ is the solution of the same equation but with continuous boundary condition.

As mentioned, the non-Hermiticity of matrix $\boldsymbol{C}$ only comes from the $\boldsymbol{C}_{rad}$ and $\boldsymbol{C}_{TE\text{-}TM}$ parts. At the symmetric $C_{4v}\text{-}\Gamma$ point, the coupling coefficients $\kappa_{m,n}=2f(1/\varepsilon_1-1/\varepsilon_2)J_1(\zeta)/\zeta, \zeta=2\sqrt{f\pi(m^2+n^2)}$ only depends on $(m^2+n^2)$. Therefore, all the non-zero elements in $\boldsymbol{C}_{rad}$ and $\boldsymbol{C}_{TE\text{-}TM}$ are identical , and hence, the imaginary part of matrix $\boldsymbol{C}$ has a symmetric form like:
\begin{equation}
\label{Eq:B.6}
\boldsymbol{B}=i\left(
\begin{array}{cccc}
d &d &0 &0\\
d &d &0 &0\\
0 &0 &d &d\\
0 &0 &d &d
\end{array}
\right)
\end{equation}

The forth part $\boldsymbol{C}_{2D}$ corresponds to the in-plane 2D distributed couplings via high-order waves. As the close form of $\boldsymbol{C}_{2D}$ is too complex, here we prove that the $\boldsymbol{C}_{2D}$ matrix also possesses a symmetric form like $\boldsymbol{A}$ in Eq.(\ref{eq:14}) by analyzing the underlying coupling paths. 

We consider the lowest set of the high-order waves that consists of four individual waves $\{(1,1),(1,-1),(-1,1),(1,1)\}$. Via these waves, there are 4 different coupling paths depicted by $\boldsymbol{C}^{2}_{2D}$. The superscript $1$ denotes that the matrix depicts the coupling via high-order waves of $m^2+n^2=2$.

\begin{figure}[!htp]
\centering
\includegraphics[width=7.2cm]{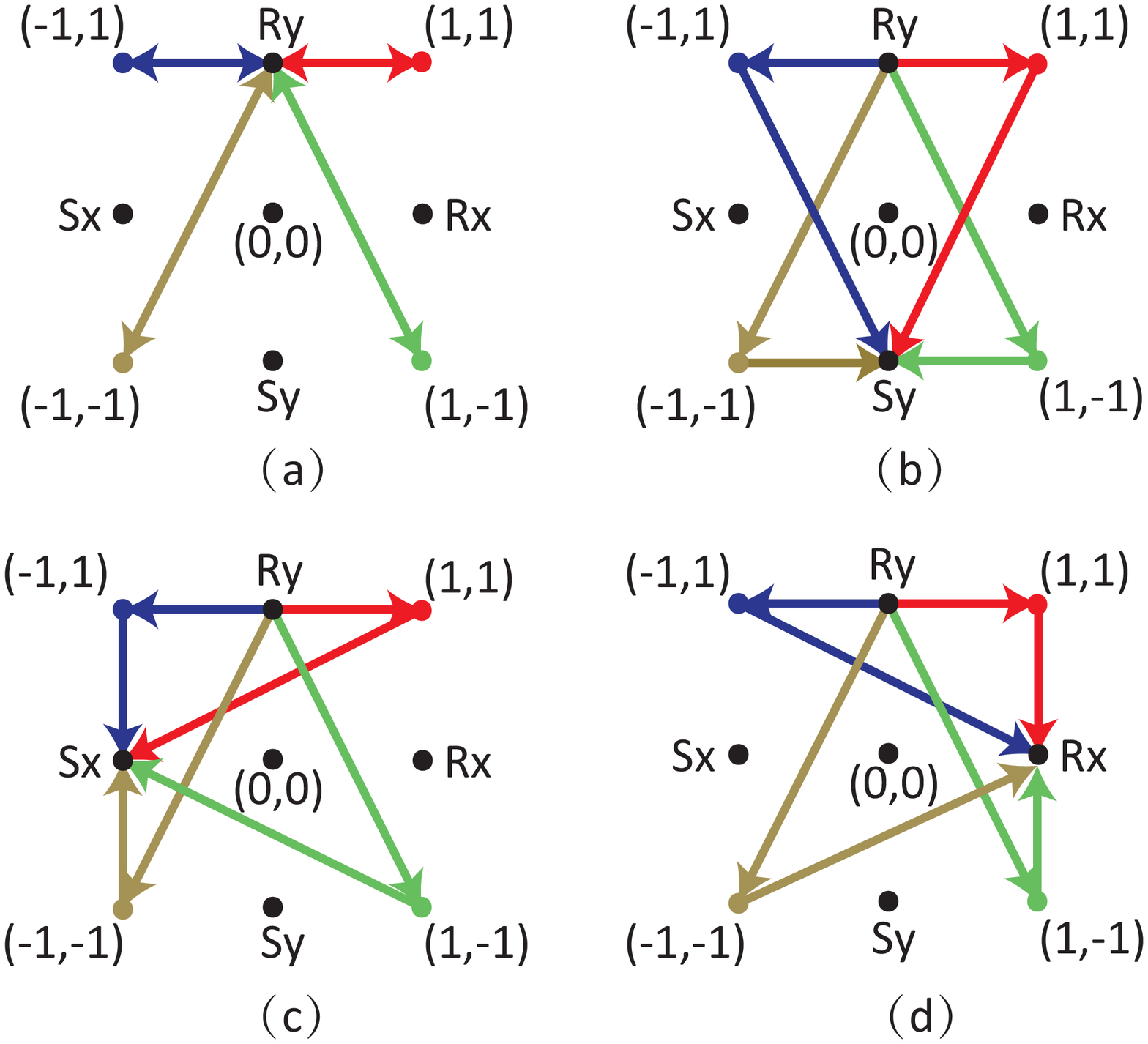}
\caption{Four possible coupling paths depicted by $\boldsymbol{C}^{2}_{2D}$ }
\label{fig:B.1}
\end{figure}

Fig. \ref{fig:B.1} (a) illustrates one possible coupling path where the basic wave $(0,1)$ couple to the four high-order waves, and then back to itself. This coupling path is depicted by the matrix elements  $\boldsymbol{C}^{2}_{2D}(3,3)$. Apparently, the four diagonal elements corresponding to this coupling path are identical.

Fig. \ref{fig:B.1} (b) presents another possible path between basic wave $(0,1)$ and $(0,-1)$. This coupling path is depicted by matrix elements $\boldsymbol{C}^{2}_{2D}(4,3)$, $\boldsymbol{C}^{2}_{2D}(3,4)$, $\boldsymbol{C}^{2}_{2D}(1,2)$ and $\boldsymbol{C}^{2}_{2D}(2,1)$.

Moreover, Fig. \ref{fig:B.1} (c) and (d) give the third and fourth possible paths between the basic waves $(0,1)$ and $(-1,0)$. These two coupling paths are depicted by matrix elements $\boldsymbol{C}^{2}_{2D}(2,3)$, $\boldsymbol{C}^{2}_{2D}(1,4)$, $\boldsymbol{C}^{2}_{2D}(3,2)$, $\boldsymbol{C}^{2}_{2D}(4,1)$,
and by $\boldsymbol{C}^{2}_{2D}(1,3)$, $\boldsymbol{C}^{2}_{2D}(2,4)$, $\boldsymbol{C}^{2}_{2D}(3,1)$, $\boldsymbol{C}^{2}_{2D}(4,2)$, respectively. Apparently, the two paths given by Fig. \ref{fig:B.1} (c) and (d) have the same coupling strength but in an opposite sign. 

Combined all the coupling paths,  the matrix $\boldsymbol{C}^{2}_{2D}$ have the a symmetric form, as:
\begin{eqnarray}
\label{eq:B.7}
\boldsymbol{A}=\left(
\begin{array}{cccc}
a &b &-c &c\\
b &a &c &-c\\
-c &c &a &b\\
c &-c &b &a
\end{array}
\right)\\ \nonumber
\end{eqnarray}

Following the same approach, we can conclude that every $\boldsymbol{C}^{M}_{2D}, M\ge 1$ has the same symmetric form with $\boldsymbol{C}^{1}_{2D}$, as a result, the entire $\boldsymbol{C}_{2D}$ owns the symmetric form as shown in Eq.(\ref{eq:B.7}). Finally, we prove that coupling matrix $\boldsymbol{C}$ has a form of Eq.(\ref{eq:14}).

\section{the matrix  $\boldsymbol{K}$ at off-$\Gamma$ points}\label{App:C}
In this section, we prove that in the vicinity of the $\Gamma$ point, the eigenvalue $k_{0,m n}, (m^2+n^2=1)$ of Eq.(\ref{eq:7}) linearly varies along $k_x$ and $k_y$. $k_{0,m n}$ is the solution to the dispersion equation \cite{39}:
\begin{equation}
\label{eq:C.1}
\tan(gt)=\frac{g\varepsilon_{\text{avg}}(\varepsilon_3p+\varepsilon_4q)}{\varepsilon_2\varepsilon_4g^2-\varepsilon_{\text{avg}}^2pq}
\end{equation}
where $t=2h$ is the thickness of PC layer. Coefficient $g, p$ and $q$ are the propagation constants of each layer:
\begin{eqnarray}
\label{eq:C.2}
g&=&\sqrt{\varepsilon_{\text{avg}}k^2-\beta^2}\nonumber\\
q&=&i\sqrt{\varepsilon_3k^2-\beta^2}\nonumber\\
p&=&i\sqrt{\varepsilon_4k^2-\beta^2}
\end{eqnarray}
where $\beta=\sqrt{m_x^2+n_y^2}\beta_0$. $k_0$ is the solution of Eq.(\ref{eq:C.1}) when $\beta=\beta_0$.

From Eq.(\ref{eq:C.2}) we know $g,p$ and $q$ all depend on $k$ and $\beta$. We concern about an arbitrary direction $n_y=n+\Delta_y, m_x=m+\Delta_x$, where $\Delta_{x,y} = k_{x,y}/\beta_0$. Expanding $g, p$ and $q$ in the vicinity of $\beta_0$, we can have
\begin{eqnarray}
\label{eq:C.3}
g&=&g_0-\frac{\beta_0^2}{g_0}(m\Delta_x+n\Delta_y)\nonumber\\
p&=&p_0+\frac{\beta_0^2}{p_0}(m\Delta_x+n\Delta_y)\nonumber\\
q&=&q_0+\frac{\beta_0^2}{q_0}(m\Delta_x+n\Delta_y)
\end{eqnarray}
where $g_0, p_0$ and $q_0$ denote the values of $g, p$ and $q$ when $\beta=\beta_0$. Expand $\tan(gt)$ in the vicinity of $\beta_0$:
\begin{eqnarray}
\label{eq:C.4}
\tan(gt)&=&\tan[t(g_0-\frac{\beta_0^2}{g_0}(m\Delta_x+n\Delta_y))]\nonumber\\
&=&\tan(g_0t)-\frac{t}{\cos^2(g_0t)}\frac{\beta_0^2}{g_0}(m\Delta_x+n\Delta_y)
\end{eqnarray}
Notice that both $\tan(g_0t)$ and $\cos^2(g_0t)$ don't depend on $\Delta_x$ or $\Delta_y$ since $g_0$ only relies on $\beta_0$. Submitting Eqs.(\ref{eq:C.3}) and (\ref{eq:C.4}) into Eq.(\ref{eq:C.1}) and neglecting the high-order terms of $(m\Delta_x+n\Delta_y)$, we obtain
\begin{widetext}
\begin{eqnarray}
\label{eq:C.5}
\tan(g_0 t)[\varepsilon_3\varepsilon_4g_0^2-\varepsilon^2_{\text{avg}}p_0q_0]=\varepsilon_{\text{avg}}g_0[\varepsilon_3p_0+\varepsilon_4q_0]+\xi \beta_0^2(m\Delta_x+n\Delta_y)
\end{eqnarray}
where
\begin{eqnarray}
\xi=\frac{t}{\cos^2(g_0t)g_0}[\varepsilon_3\varepsilon_4g_0^2-\varepsilon^2_{\text{avg}}p_0q_0]+\tan(g_0 t)[2\varepsilon_3\varepsilon_4+\varepsilon_{\text{avg}}^2(\frac{p_0}{q_0}+\frac{q_0}{p_0})]+\varepsilon_{\text{avg}}[\varepsilon_3(\frac{g_0}{p_0}-\frac{p_0}{g_0})+\varepsilon_4(\frac{g_0}{q_0}-\frac{q_0}{g_0})]\nonumber
\end{eqnarray}
\end{widetext}

$g_0, p_0, q_0$ and $\xi$ in Eq.(\ref{eq:C.5}) only rely on  $\beta_0$ since their dependences to $(m\Delta_x+n\Delta_y)$ have been separated into the linear terms. All of them are functions of unknown quantity $k$ and we can expand them near zero-order solution $k_0$ as
\begin{eqnarray}
\label{eq:C.6}
g_0&=&g_{00}+\frac{\varepsilon_{\text{avg}}\sigma}{g_{00}}k_0\nonumber\\
p_0&=&p_{00}-\frac{\varepsilon_{4}\sigma}{g_{00}}k_0\nonumber\\
q_0&=&q_{00}-\frac{\varepsilon_{3}\sigma}{q_{00}}k_0
\end{eqnarray}
where $\sigma=k-k_0$ and $g_{00}, p_{00}$ and $q_{00}$ denote the values of $g, p$ and $q$ when $\beta=\beta_0, k=k_0$. Since $\xi$ is a slow varying function with respect to $k$, we approximately take $\xi=\xi_0(k_0)$ which can be evaluated through the structural parameters and $k_0$ directly. 

Expand $\tan(gt)$ in the vicinity of $k_0$ similarly:
\begin{equation}
\label{eq:C.7}
\tan(g_0 t)=\tan(g_{00}t)+\frac{1}{\cos^2(g_{00}t)}\frac{\varepsilon_{\text{avg}}k_0}{g_{00}}\sigma
\end{equation}

Submitting Eqs.(\ref{eq:C.6}) and (\ref{eq:C.7}) to Eq.(\ref{eq:C.5}), we have
\begin{widetext}
\begin{eqnarray}
\label{eq:C.8}
\tan(g_{00} t)[\varepsilon_3\varepsilon_4g_{00}^2-\varepsilon^2_{\text{avg}}p_{00}q_{00}]=[\varepsilon_{\text{avg}}\varepsilon_3g_{00}p_{00}+\varepsilon_{\text{avg}}\varepsilon_4g_{00}q_{00}]+\varsigma k_0\sigma+\xi\beta_0^2(m\Delta_x+n\Delta_y)
\end{eqnarray}
where
\begin{eqnarray}
\varsigma=\varepsilon_{\text{avg}}[\varepsilon_{\text{avg}}(\frac{\varepsilon_3 p_{00}}{g_{00}}+\frac{\varepsilon_4 q_{00}}{g_{00}})-\varepsilon_3\varepsilon_4(\frac{g_{00}}{p_{00}}+\frac{g_{00}}{q_{00}})]-\tan(g_{00} t)[2\varepsilon_3\varepsilon_4\varepsilon_{\text{avg}}+\varepsilon^2_{\text{avg}}(\frac{\varepsilon_3 p_{00}}{q_{00}}+\frac{\varepsilon_4 q_{00}}{p_{00}})]-\frac{[\varepsilon_3\varepsilon_4g^2_{00}-\varepsilon^2_{\text{avg}}p_{00}q_{00}]}{\cos^2(g_{00} t)}\frac{\varepsilon_{\text{avg}}}{g_{00}}\nonumber
\end{eqnarray}
\end{widetext}
Apparently, Eq.(\ref{eq:C.8}) can be simplified as $\varsigma k_0\sigma+\xi\beta_0^2(m\Delta_x+n\Delta_y)=0$, that is
\begin{equation}
\label{eq:C.9}
k=k_0+\eta(m\Delta_x+n\Delta_y)
\end{equation}
where $\eta=-\xi \beta_0^2/(\varsigma k_0)$.

$\xi=\xi_0(k_0)$ and $\varsigma$ all are determined by to $\beta_0$ and $k_0$, which can be evaluated by the zero-order trial solution and structural parameters. As $\boldsymbol{k}_T=(k_x\hat{e_x}+k_y\hat{e_y})=\beta_0(\Delta_x\hat{e_x}+\Delta_ y\hat{e_y})$, Eq.(\ref{eq:C.9}) turns to be Eq.(\ref{eq:21}).

It is noteworthy that,  the coefficient $\eta$ depends on a series of structural parameters. It becomes tedious to analytically evaluate $\eta$ according to Eq.(\ref{eq:C.5}) and (\ref{eq:C.8}). Alternatively, we can obtain $\eta$ by fitting the dispersion curve given by Eq.(\ref{eq:C.1}).   

\bibliography{DIrac}

\end{document}